\documentclass[galaxies,conferencereport,accept,moreauthors,pdftex,10pt,a4paper]{mdpi}

\firstpage{1}
\articlenumber{x}
\doinum{10.3390/------}
\pubvolume{xx}
\pubyear{2018}
\copyrightyear{2018}
\history{Received: 31 January 2018; Accepted: 22 February 2018; Published: 25 February 2018}

\pdfoutput=1

\usepackage{units}
\newcommand{\fe}{$^{60}\text{Fe}$}

\usepackage{booktabs}
\usepackage{multirow}
\usepackage{soul}
\usepackage{microtype}

\Title{A Way Out of the Bubble Trouble?---Upon Reconstructing the Origin of the Local Bubble and Loop~I via Radioisotopic Signatures on Earth}

\Author{\mbox{Michael Mathias Schulreich $^{1,}$*\orcidA{},
~Dieter Breitschwerdt $^{1}$, Jenny Feige $^{1}$ and Christian Dettbarn $^{2}$}}
\AuthorNames{Michael Mathias Schulreich, Dieter Breitschwerdt, Jenny Feige, and Christian Dettbarn}

\address{%
$^{1}$ \quad Zentrum f\"{u}r Astronomie und Astrophysik, Technische Universit\"{a}t Berlin, Hardenbergstra{\ss}e 36, 10623~Berlin,~Germany; breitschwerdt@astro.physik.tu-berlin.de (D.B.); feige@astro.physik.tu-berlin.de (J.F.) \\
$^{2}$ \quad \mbox{\textls[-15]{Astronomisches Rechen-Institut, Zentrum f\"ur Astronomie der Universit\"at Heidelberg, M\"onchhofstra{\ss}e 12--14,}} 69120 Heidelberg, Germany; dettbarn@ari.uni-heidelberg.de}

\corres{Correspondence: schulreich@astro.physik.tu-berlin.de}
\abstract{Deep-sea archives all over the world show an enhanced concentration of the radionuclide {\fe}, isolated in layers dating from about $\unit[2.2]{Myr}$ ago. Since this comparatively long-lived isotope is not naturally produced on Earth, such an enhancement can only be attributed to extraterrestrial sources, particularly one or several nearby supernovae in the recent past. It has been speculated that these supernovae might have been involved in the formation of the Local Superbubble, \mbox{\textls[-15]{our~Galactic habitat. Here, we summarize our efforts in giving a quantitative evidence for this scenario.}} Besides~analytical calculations, we present results from high-resolution hydrodynamical simulations of the Local Superbubble and its presumptive neighbor Loop~I in different environments, including a self-consistently evolved supernova-driven interstellar medium. For the superbubble modeling, the time sequence and locations of the generating core-collapse supernova explosions are taken into~account, which are derived from the mass spectrum of the perished members of certain, carefully~preselected stellar moving groups. The release and turbulent mixing of {\fe} is followed via passive scalars, where the yields of the decaying radioisotope were adjusted according to recent stellar evolution calculations. The models are able to reproduce both the timing and the intensity of the {\fe} excess observed with rather high precision. We close with a discussion of recent developments and give future perspectives.
}


\begin{document}
%%%%%%%%%%%%%%%%%%%%%%%%%%%%%%%%%%%%%%%%%%
%% Only for the journal Gels: Please place the Experimental Section after the Conclusions

%%%%%%%%%%%%%%%%%%%%%%%%%%%%%%%%%%%%%%%%%%
%\setcounter{section}{-1} %% Remove this when starting to work on the template.
\section{Introduction}
\label{ch:intro}
Superbubbles (SBs) are gigantic structures carved into the interstellar medium (ISM) by fast stellar winds and supernova (SN) explosions within groups of massive stars. Observational characteristics are a shell of neutral gas with several hundreds of parsecs to even a few kiloparsecs radius, enclosing~a soft X-ray emitting volume of hot, low-density gas. Remarkably, our solar system seems to reside in such an~environment, which led to the conceptualization of the Local (Hot) Bubble (LB) \cite{San:77,Inn:84,Bre:94,Smi:01}. \mbox{\textls[-22]{Also~embedded in this region is the Local Interstellar Cloud (LIC)---a structure of slightly higher density,}} with a diameter of about $\unit[5]{pc}$. The Sun and its heliosphere (about 200--$\unit[300]{au}$ in size) entered the LIC within the past $10^4$ years and are currently located either still at the edge of it or already within the transition zone between the LIC and another cloud referred to as the G cloud \cite{Fri:11}.

Linking the LB's existence to a specific stellar aggregate that has lost a fraction of its most massive members in SN explosions is not a trivial task, as currently, the LB is devoid of suitable candidates. It was therefore obvious to look for a moving group of stars that had passed through the present LB volume somewhat earlier in time \cite{Ber:02,Mai:01,Fuc:06}. In Fuchs et al.~(2006) \cite{Fuc:06}, B stars within a heliocentric sphere of $\unit[400]{pc}$ diameter were selected according to their compactness in real and velocity space from astrometric catalogue data (such as Hipparcos), and then traced backwards in time by solving the epicyclic equations of motion. As an update to this approach, the \emph{most probable} center-of-mass trajectories were calculated instead, which result from also taking into account Gaussian errors in the Hipparcos positions and proper motions \cite{Bre:16}. It was found that the cluster had entered the current LB region rather off-center about \unit[13]{Myr} ago. As all the stars in the cluster were likely born at the same~time, the cluster's age coincides with the core hydrogen-fusion lifetime of the hottest, most~luminous stars. The position of the main-sequence turn-off point, which is an estimator for that, was determined via isochrone fitting, yielding a cluster age between 20 and $\unit[30]{Myr}$.~Assuming that the stellar mass spectrum is represented by \mbox{an initial mass function (IMF) typical for young massive stars~\cite{Mas:95},} with~only one star per mass bin---for the distribution with the least statistical bias ---16 stars with masses ranging from 8.81 to $\unit[19.86]{M_\odot}$
~were obtained. Their main-sequence lifetimes were calculated as a function of mass from the isochrones \cite{Sch:92}. Subtracting the cluster age from these lifetimes yielded the SN explosion times, which had then to be combined with the most probable stellar trajectories to give the most probable explosion sites. The surviving cluster members belong to the Sco-Cen~association.

Further information on the close SNe that shaped the LB can be gained from looking at their possible traces on Earth. Certain radioisotopes that are uniquely associated with SN activity soon crystallized out as potential targets for such astroarcheological endeavors, requiring that they live long enough not only  to travel significant distances (a few hundred parsecs) through the ISM, but~also to remain detectable till the present day, after their arrival on our planet. Listed as a particularly promising candidate, already by these early estimates  \cite{Kor:96,Ell:96}, is {\fe}, which is produced in the Ne--O zones of massive stars, both hydrostatically and explosively, by successive neutron capture on $^{58}$Fe and $^{59}$Fe, and, furthermore, in the base of the He-shell, by a less-vigorous r-process during the explosion~\cite{Woo:95}. Its modern half-life is about \unit[2.6]{Myr} \cite{Rug:09,Wal:15}. After its release, its $\beta^{-}$ decay via $^{60}$Co becomes directly measurable as diffuse Galactic $\gamma$-ray emission at 1173 and $\unit[1333]{keV}$ \cite{Wan:07}, proving~the rather widespread, ongoing nucleosynthesis in our Milky Way. Without recent SN activity in our~vicinity, {\fe} should not occur on Earth because of its non-existent natural terrestrial production, its negligible extraterrestrial influx via interstellar dust particles and micrometeorites, and the fact that all primordial {\fe} has had plenty of time to decay. Therefore, it was all the more surprising when in 1999 live {\fe} was found inside two layers of a ferromanganese (FeMn) crust from the South Pacific Ocean floor, dating back about $\unit[6]{Myr}$ \cite{Kni:99}. A characteristic of these crusts, which they by the way share with slow-accumulating deep-sea sediments and nodules, is their remote position, allowing for a particularly uniform and undisturbed growth via the incorporation of chemical elements. The growth rate of crusts and nodules is only about a few millimeters and centimeters per million years, respectively, and is due to precipitation of manganese and iron---the main constituents of these archives. Sediments, on the other hand, are formed by the deposition of particles that originate mainly from the surface of the sea, such as the remains of deceased organisms or continental dust. Even particles from the sea itself, for example, from submarine volcanic eruptions, can get into the sediments. This is why they grow more than a thousand times faster than FeMn crusts, thus offering a better time resolution (cf.~\cite{Fei:16}). A~deeper analysis of another FeMn crust sample (termed 237KD)---this time from the equatorial Pacific---was published in 2004 \cite{Kni:04}. In this study, which, like the former one, made use of accelerator mass spectrometry, the crust sample was sliced into up to $\unit[2]{mm}$ thick layers, corresponding to an age range of up to $\unit[800]{kyr}$ per layer. Owing to this detailed dissection it was possible to detect an unambiguous {\fe} anomaly in a segment that is between 1.7 and 2.6 Myr old. Four years later, the~signal inside 237KD was confirmed by another research group \cite{Fit:08}.

We set ourselves the goal to explain these measurements within the framework of the above described LB formation scenario. To that end, we first performed analytical calculations, where we considered, in contrast to other studies (e.g., \cite{Fry:15}), not a single SN event, but the aforementioned \mbox{\textls[-10]{sequence of explosions \cite{Fei:10,Bre:16}. This was implemented using a SN model introduced by F.~D.~Kahn \cite{Kah:98},}} which delivers the expansion of SN remnants in an external medium already stratified by an earlier~explosion. Additionally, we required that the blast waves never overtook the also-expanding outer shell of the LB (treated following Weaver et al.~(1977) \cite{Wea:77}), since sequentially propagating shock waves always coalesce. The mass of the ejecta---including the proportion of {\fe}---was taken, according to the mass of the progenitor, from fitting stellar evolution calculations \cite{Woo:95,Rau:02,Lim:06,Woo:07}.

Although the results obtained with this rather simple approach already show surprisingly good agreement with the crust data---when applying a canonical {\fe} survival fraction (see~Section~\ref{ch:nummod})---they are still based on simplifying assumptions (zero external pressure, time-frozen homogeneous or power-law-like external density distribution, negligence of turbulent mixing or mass loading), which~can only be dropped by numerically solving the full-blown set of the governing fluid-dynamical~equations. This will be discussed in the next section.

\section{Numerically Modeling the $^{60}$Fe Transport to the Solar System}
\label{ch:nummod}

\begin{figure}[H]
\centering
\includegraphics[width=12cm]{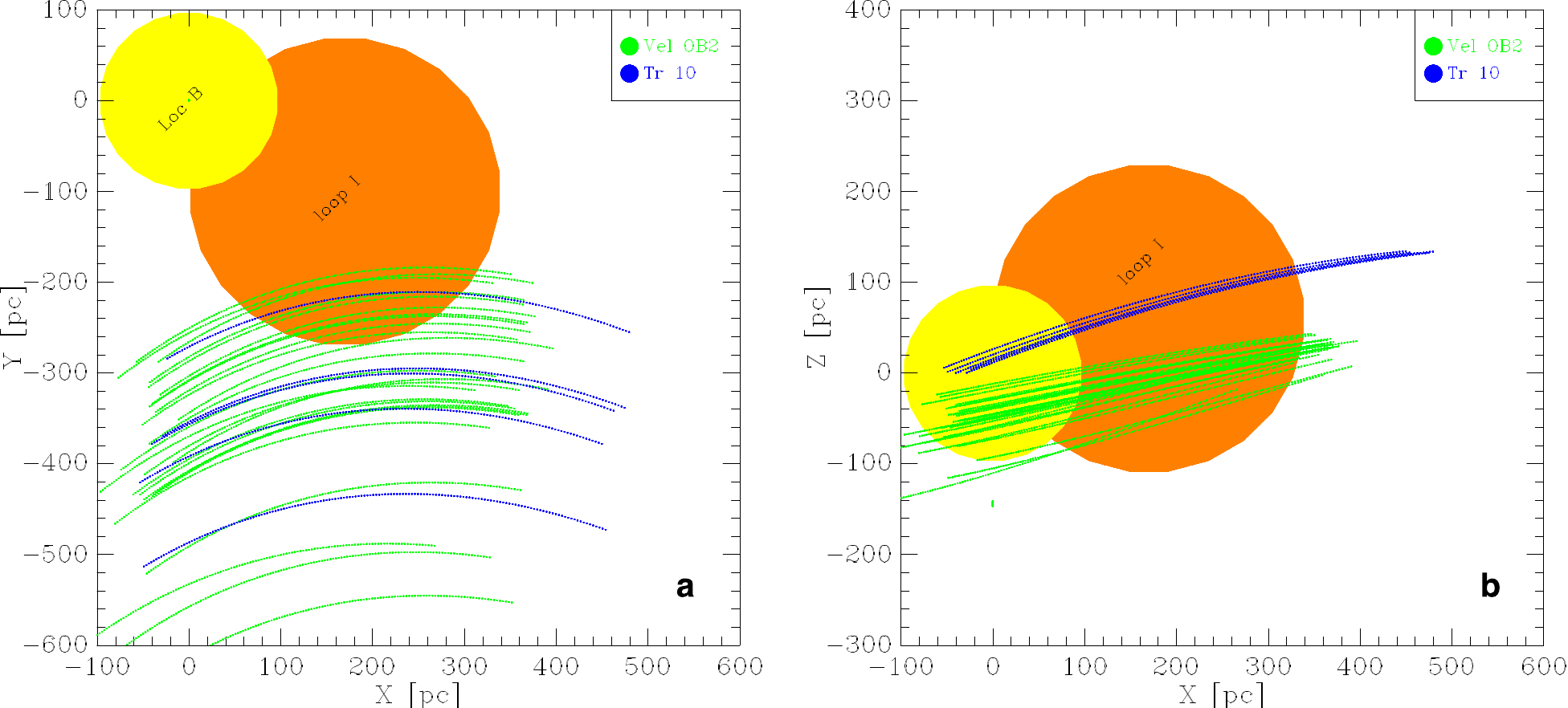}
\caption{Calculated possible trajectories of the Loop~I progenitor stars associated with Vel~OB2 (green~dots) and Tr~10 (blue dots), plotted together with the supposed current Local and Loop~I superbubble region in projection onto a slice (\textbf{a}) along and (\textbf{b}) perpendicular to the Galactic midplane. Our solar system resides at the coordinate origin.}
\label{im:loopi}
\end{figure}

Like in a similar study by Breitschwerdt \& de Avillez (2006) \cite{Bre:06}, we considered the evolution of our LB not in isolation but together with its neighbor, the Loop~I SB. Recently there have been some doubts in the literature (see, e.g., \cite{Pus:14}) whether Loop~I and the North Polar Spur, one of the best-known features in radio continuum and diffuse soft X-ray maps, which is speculated to be associated with Loop~I, is part of the local ISM at all or rather lies closer to the Galactic center. In view of the numerous nearby young stars in that direction, it would however be quite surprising if there was no SB in our direct neighborhood at all. We actually made use of that fact when we set up Loop~I as a ``boundary condition'' for our model: repeating in principle the procedure described in Section~\ref{ch:intro} (for details also see \cite{MMS:17}), we namely searched for clusters of B stars within a doubled spherical volume ($\unit[800]{pc}$ diameter) in order to identify possible SN progenitors. We found that Tr~10 and the association Vel~OB2 have recently crossed the presumptive present-day Loop~I region and that they have probably witnessed 19 SN explosions. The trajectories along which these SNe might have taken place are shown in Figure~\ref{im:loopi}.~These were again derived from solving the epicyclic equations. Note that in all the maps depicted in this work the $x$, $y$, and $z$-axes point toward the Galactic center, into the direction of the Galactic rotation, and toward the Galactic north pole, respectively. Since, as can be seen, not all trajectories pass through Loop~I, we selected those that lie closest to its center. The interaction between the two SBs also promotes the generation of a neutral gas wall, which has been observed in X-ray absorption \cite{Egg:95}.

\begin{figure}[H]
\centering
\includegraphics[width=15cm]{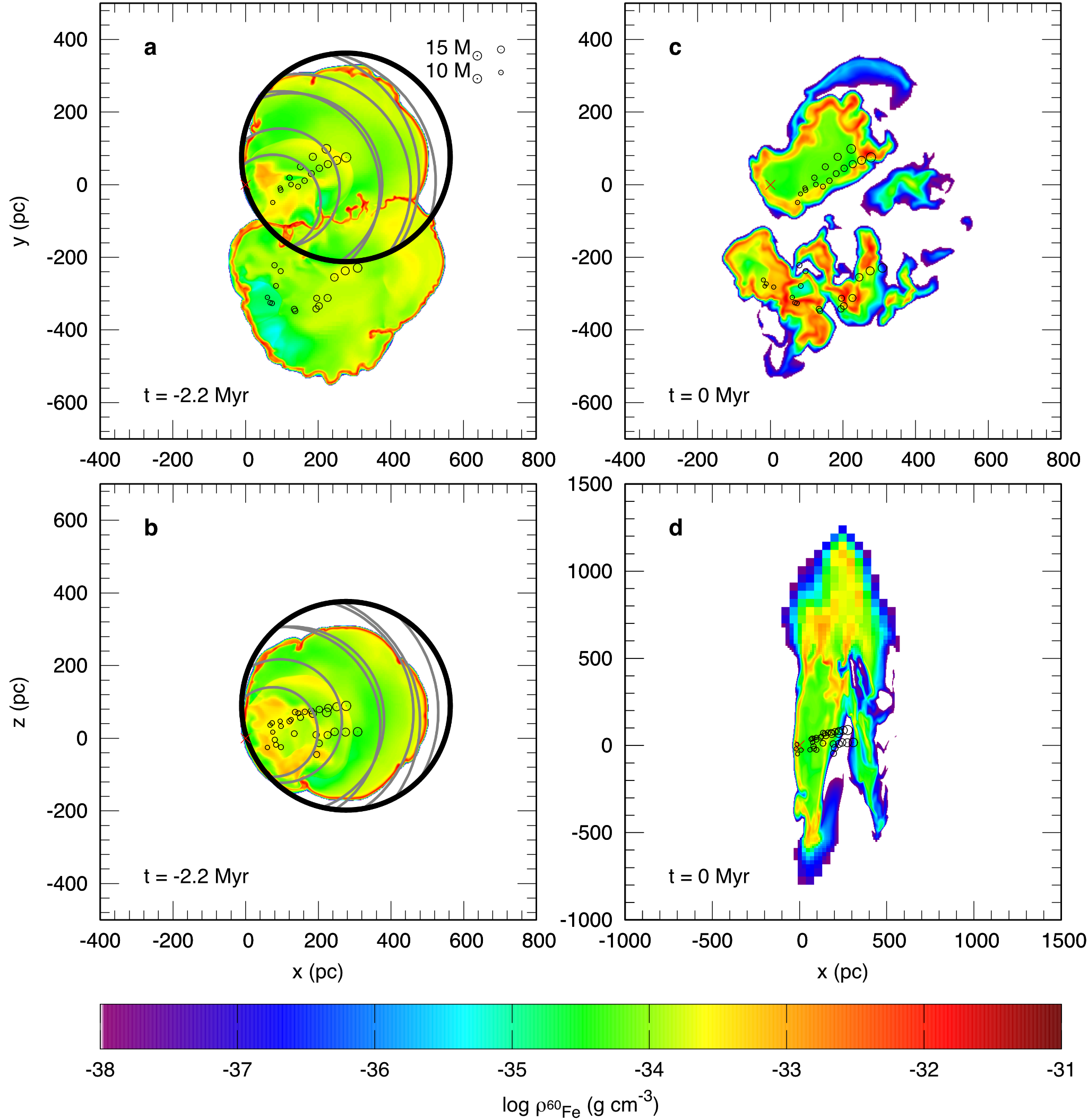}
\caption{Color-coded maps of the logarithmic {\fe} mass density distribution associated with the Local and Loop~I superbubble, as derived from the numerical simulations with (\textbf{a}, \textbf{b}) homogeneous background medium ($\unit[0.3]{cm^{-3}}$ number density, $\unit[6800]{K}$ temperature, solar metallicity) at $\unit[2.2]{Myr}$ before present and (\textbf{c}, \textbf{d}) inhomogeneous background medium as at present day. Vertical ($y=0$) and horizontal ($z=0$) cuts through the three-dimensional computational box are shown in the panels \textbf{a},~\textbf{c} and \textbf{b}, \textbf{d}, respectively. Earth's projected position is at $(0,0)$ and marked by a red cross. Thin black circles indicate the projected centers of the supernova explosions that occurred in the timeframe before the snapshot was taken. The sizes of the circles correspond to the initial masses of the progenitor stars (see legend in panel \textbf{a}). Overplotted in the panels \textbf{a} and \textbf{b} are the results of the analytical calculations, with the Local Bubble's outer shell depicted as a thick black circle, and the individual supernova blast waves in its interior as gray curves.}
\label{im:fedmaps}
\end{figure}

Now that we had pinned down for each of the two SBs the number of SN progenitors, their~explosion sites, their initial masses (via an IMF), together with their explosion times, total ejected masses, and {\fe} mass fractions (via stellar evolution models), we could simulate the evolution of the SNe in three dimensions using the finite-volume shock-capturing scheme RAMSES \cite{Tey:02} that is based on a second-order extension of Godunov's method for solving hyperbolic systems of conservation laws (see, e.g., \cite{Tor:09}). Owing to its (tree-based) adaptive mesh refinement technique, numerical resolutions down to subparsec scale could be achieved. We launched the explosions into two different kinds of environments. On the one hand, homogeneous self-gravitating media that feature conditions covered by the classical three-phase model \cite{Kee:77}, and, on the other hand, a medium designed to mimic more realistic conditions in the local Galaxy. These are achieved by exposing an initial interstellar gas distribution, derived from observations, for $\unit[180]{Myr}$ to the combined effects of the Galactic gravitational field, various heating and cooling processes, winds of massive stars (which are numerically treated as particles forming at Galactic rate in cold and dense interstellar clouds), and SNe. Gas in the resulting Galactic disk expresses typical features of a compressible medium subject to supersonic explosion-driven turbulence, where also the thermal properties are found to be in satisfactory agreement with observations and predictions from similar models (e.g, \cite{Avi:04}). The dynamics of the decaying {\fe} was followed via so-called passive scalars or tracers, which are quantities that behave like a drop of ink when dispersed in a liquid. Figure~\ref{im:fedmaps} displays {\fe} mass density distributions associated with the two SBs (in the panels a and c, the LB is at the top) for both kinds of background~environments. It can be seen that the SBs embedded into the inhomogeneous background medium are generally much more irregular and vertically elongated, as a result of steep density and pressure gradients perpendicular to the Galactic midplane. The highest {\fe} concentrations are found in the supershells as well as in the shells of the individual SN remnants. For the homogeneous background medium case (panels a and b) we also show, for comparison, the results of the analytical~calculations. These generally overestimate the size of the LB, even in directions where its does not interact with Loop~I, which is mainly due to their total neglect of the external pressure.

\begin{figure}[H]
\centering
\includegraphics[width=13cm]{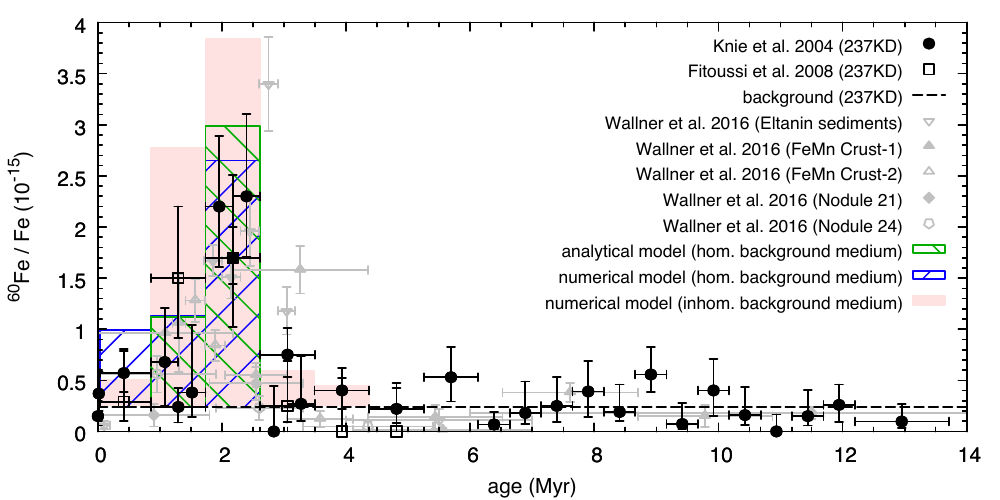}
\caption{Comparison of the measured (symbols with error bars) and calculated (histograms) {\fe}/Fe ratio as a function of the terrestrial archive's layer age. Modeled data is set upon the instrumental background derived specifically for deep-sea crust sample 237KD, which is indicated by the dashed~line.}
\label{im:vglcrust}
\end{figure}

By tagging the {\fe}-enriched gas -- not only as a whole but also the contributions of each individual SN---we measured the flux of {\fe} atoms at the Earth's position over the entire simulation time, and~then smeared out this so-called fluence into bins corresponding to the time resolution of the FeMn crust sample 237KD. For a direct comparison with the measurements, however, one also has to take into account that only a small fraction of the {\fe} atoms is actually able to reach the Earth's orbit after overcoming a variety of filtering processes (see, e.g., \cite{Fry:15}), while being incorporated in dust grains---otherwise it would not be able to propagate against the heliospheric ram pressure. In~addition, possibly not all the {\fe} dust spread over Earth's surface finds its way into the FeMn crust, as a result of chemical selection processes. In accordance with earlier studies \cite{Kni:04}, we estimated a {\fe} survival fraction of 0.006, with which we had to multiply our calculated fluences. The resulting profiles for our best-fitting numerical and analytical models are shown, together with all currently available deep-sea~measurements, in Figure~\ref{im:vglcrust}. Although in all cases the timing and intensity of the observed {\fe} excess is well reproduced, the underlying physical processes are different. In the inhomogeneous case (pale red histogram), the maximum signal is due to two individual SN events (number 14 and 15 out of the total of 16, at a distance of $\sim$106 and $\unit[91]{pc}$, respectively), which cross Earth's orbit twice as a result of shock reflection from the LB's outer shell. In the two presented models with homogeneous background medium (numerical one in blue; analytical one in green), it is the supershell of the LB itself that delivers the live {\fe} content of all previous SNe (numbers 1 to 15) at once, thereby stretching the deposition over several hundred thousand years. The later arriving blast wave of the 16th SN, \mbox{\textls[-23]{which~occurred at a distance of $\sim$$\unit[96]{pc}$ to the solar system, again left isotopic marks on Earth $\unit[1.5]{Myr}$ ago.}} Some sort of ``{\fe} background noise'' is produced through turbulent motions inside the SB cavity, introduced by asymmetrically reflected shock waves. From all models run we found that the density of the ambient medium must not exceed $\unit[0.3]{cm^{-3}}$, on average. Otherwise, the LB supershell would arrive too late for the {\fe} overabundance to occur in the crust layer in which it was discovered.

\unskip

\section{Discussion and Outlook}
At the same time when we first published these results, a paper appeared reporting that the {\fe} signal has now been measured in deep-sea archives from all major oceans, proving that it is indeed a global phenomenon \cite{Wal:16}. Furthermore, because of the better time resolution of sediments, studied~alongside several FeMn crusts and nodules, it was found that the peak is actually broader (1.5--\unit[3.2]{Myr} ago) than indicated by the previous measurements. Both is compatible with our multi-SN LB formation scenario, as can be seen from the gray data points in Figure~\ref{im:vglcrust}.~Also noteworthy is the occurrence of an additional smaller peak 6.5--$\unit[8.7]{Myr}$ ago, ``whose exact origin is yet rather elusive. Remarkably, a collision of asteroids in the main belt $\unit[8.3]{Myr}$ ago, connected to a boosted, possibly {\fe} enriched, influx of interplanetary dust particles and micrometeorites \cite{Far:06} falls within this particular time range''~\cite{MMS:17}. Meanwhile, several other findings related to recent SN activity in the solar neighborhood have emerged:
\begin{enumerate}[leftmargin=10mm,labelsep=3mm]
\item	Measurements in two independent Pacific Ocean sediment cores revealed elevated {\fe} levels in microfossils dated at 1.8--$\unit[2.6]{Myr}$ \cite{Lud:16}. These are the remains of so-called magnetotactic bacteria, which feed on iron to produce chains of magnetite (Fe$_3$O$_4$) crystals (so-called magnetosomes) for orientation at Earth's magnetic field. When the bacteria population moves upward as the sediment grows, microfossils are left behind and the magnetite crystals get preserved in the corresponding sediment layers.
\item	Enhanced {\fe} signatures were detected  in lunar soil samples recovered during the Apollo missions 12, 15, and 16 \cite{Fim:16}. Unfortunately, the almost atmosphere-free Moon allows for no time-resolved measurements due to layer-mixing as a result of the continuous meteoritic bombardment (``gardening''). Solar and galactic cosmic rays (CRs) can also generate {\fe} (and~$^{53}$Mn)---however, their contribution is less than 10\,\% so that the bulk of {\fe} should be from SNe. We have shown in \cite{MMS:17} that the lower limit of the detected integrated fluence ($\unit[10^{7}]{at\,cm^{-2}}$) is compatible with our LB model.
\item	The ACE-CRIS experiment detected 15 {\fe} atoms from a total of $3.55\times 10^5$ CR particles during the time period 1997--2014 \cite{Bin:16}. Since the CRIS energy range is $\sim$100 to $\unit[500]{MeV/nuc}$, acceleration of nuclei must have been due to SN blast waves (i.e., first-oder Fermi process). By~comparing the measured {\fe}/$^{56}$Fe ratio with results from stellar evolution models, the~authors concluded that the time between nucleosynthesis and acceleration is a few million years. Using a diffusive propagation model as a basis, the mean lifetime of {\fe} dictates the distance to the source to be less than $\unit[620]{pc}$, which is easily fulfilled even by the farthest explosion in our LB model ($\unit[300]{pc}$). Moreover, it was shown that the whole set of known peculiar features of the locally observed CR spectrum can be explained in the framework of a single self-consistent model including the contribution to the CR flux of a SN that has injected CRs within a distance of about $\unit[100]{pc}$ from the Sun some 2--$\unit[3]{Myr}$ ago \cite{Kac:17}. Also this is consistent with the SNe derived from our model.
\item	Three-dimensional maps of the local interstellar dust, based on the inversion of color excess measurements for individual target stars, diffuse interstellar bands, and statistical methods using stellar surveys (including Gaia), showed four soft X-ray emitting cavities that are open toward the Sun. Strikingly, two of these potential SN relics match the sites of the two most recent SNe estimated in our LB model with respect to both distance and direction (within 3$^\circ$ and 7$^\circ$, respectively) \cite{Cap:17}.
\item There is some speculation as to whether the Tuc-Hor association rather than Sco-Cen could have hosted the SN(e) responsible for the {\fe} signal \cite{Mam:16}. The masses of the current Tuc-Hor members as well as the fact that the group was at a similar distance ($\sim$$\unit[60]{pc}$) $\unit[2.2]{Myr}$ ago indeed allow for this possibility. A suggestion by Fry et al.~(2016) \cite{Fry:16} to use the Moon as an ``antenna'' for pinning down the direction of incidence of {\fe} dust and thus the responsible stellar group, however, fails since the {\fe} concentrations of soil samples, scattered widely across the lunar surface, are barely different. Due to the huge spatial extent of Tuc-Hor, which appears to be rather an ensemble of evaporating subgroups than one large group \cite{Mam:16}, it could not be captured by our selection criterion based on compactness in both real and velocity space---one could say that in this case we did not see the forest for the trees. We plan to expand future model calculations in this regard.
\item	Provided that there had been one or more near-Earth SNe 2--$\unit[3]{Myr}$ ago, it is quite likely that such an extremely bright stellar event---visible even in daylight---would had arisen the attention of the \emph{Australopithecus}. Thomas et al.~(2016) \cite{Tho:16} modeled the impact of a SN occurring at a distance of $\unit[100]{pc}$ on the terrestrial atmosphere and biota. They found that it would only have ``a small effect on terrestrial organisms from visible light and that chemical changes such as ozone depletion are weak. However, tropospheric ionization right down to the ground, due to the penetration of $\ge$TeV CRs, will increase by nearly an order of magnitude for thousands of years, and~irradiation by muons on the ground and in the upper ocean will increase twentyfold, which~will approximately triple the overall radiation load on terrestrial organisms. Such~irradiation has been linked to possible changes in climate and increased cancer and mutation rates. This may be related to a minor mass extinction around the Pliocene-Pleistocene boundary'' \cite{Tho:16} about $\unit[2.5]{Myr}$ ago.
\end{enumerate}

Fortunately, our currently pure hydrodynamical model---possible magnetic field effects are planned to be considered next---comes with hardly any fine-tuning, which is particularly demonstrated by the choice of an unbiased IMF mass binning and a representative background-ISM. It is however questionable whether a specific problem, like the formation of the LB, with all its observable details, is~actually fully ascertainable via such an unspecific approach. For further constraining our initial and boundary conditions, we therefore expect a lot of insights from the Gaia astrometric data releases in the near future.

\vspace{6pt}

%%%%%%%%%%%%%%%%%%%%%%%%%%%%%%%%%%%%%%%%%%
\acknowledgments{M.M.S. and D.B. acknowledge funding by the DFG priority program 1573 ``Physics of the Interstellar Medium''. We also thank the two anonymous referees for their valuable comments to improve the quality of the manuscript.}

%%%%%%%%%%%%%%%%%%%%%%%%%%%%%%%%%%%%%%%%%%
\authorcontributions{M.M.S. performed extensive numerical simulations on the background ISM, the evolution of the Local and Loop~I SB and the {\fe} transport, produced Figures~\ref{im:fedmaps} and \ref{im:vglcrust}, and wrote the paper. D.B. worked out the original model and led the research. J.F. carried out analytical calculations, interpreted the crust data, calculated the IMF, and contributed to Figures~\ref{im:fedmaps} and \ref{im:vglcrust}. C.D. performed the analysis of the moving group stars, calculated the trajectories of both LB and Loop~I progenitor stars, and produced Figure~\ref{im:loopi}.}

%%%%%%%%%%%%%%%%%%%%%%%%%%%%%%%%%%%%%%%%%%
\conflictsofinterest{The authors declare no conflict of interest.}

%%%%%%%%%%%%%%%%%%%%%%%%%%%%%%%%%%%%%%%%%%
%% optional
\abbreviations{The following abbreviations are used in this manuscript:\\

\noindent
\begin{tabular}{@{}ll}
CR  & cosmic ray\\
FeMn & ferromanganese\\
IMF & initial mass function\\
ISM & interstellar medium\\
LB & Local Bubble\\
LIC & Local Interstellar Cloud\\
SB & superbubble\\
SN & supernova
\end{tabular}}


%=====================================
% References, variant A: internal bibliography
%=====================================
\reftitle{References}

%\begin{thebibliography}{999}
% Reference 1
%\bibitem[Author1(year)]{ref-journal}
%Author1, T. The title of the cited article. {\em Journal Abbreviation} {\bf 2008}, {\em 10}, 142-149, DOI.
% Reference 2
%\bibitem[Author2(year)]{ref-book}
%Author2, L. The title of the cited contribution. In {\em The Book Title}; Editor1, F., Editor2, A., Eds.; Publishing House: City, Country, 2007; pp. 32-58, ISBN.
%\end{thebibliography}

% The following MDPI journals use author-date citation: Arts, Econometrics, Economies, Genealogy, Humanities, IJFS, JRFM, Laws, Religions, Risks, Social Sciences. For those journals, please follow the formatting guidelines on http://www.mdpi.com/authors/references
% To cite two works by the same author: \citeauthor{ref-journal-1a} (\citeyear{ref-journal-1a}, \citeyear{ref-journal-1b}). This produces: Whittaker (1967, 1975)
% To cite two works by the same author with specific pages: \citeauthor{ref-journal-3a} (\citeyear{ref-journal-3a}, p. 328; \citeyear{ref-journal-3b}, p.475). This produces: Wong (1999, p. 328; 2000, p. 475)

%=====================================
% References, variant B: external bibliography
%=====================================
%\externalbibliography{yes}
%\bibliography{bubbletrouble}

\end{document}